\documentclass[]{aa}
\usepackage{graphicx,txfonts,amssymb,natbib}
\renewcommand{\d}{\mathrm{d}}
\authorrunning{C. Fedeli, et al.}
\titlerunning
  {Halo concentration distribution}

\begin{document}

\title
  {Effects of the halo concentration distribution on strong-lensing optical
    depth and X-ray emission}

\author{C. Fedeli\inst{1}\thanks{E-mail: cosimo@ita.uni-heidelberg.de}, 
  M. Bartelmann\inst{1}, M. Meneghetti\inst{2} and
  L. Moscardini\inst{3,4}}
\institute{$^1$ Zentrum f\"ur Astronomie, ITA, Universit\"at Heidelberg,
     Albert-\"Uberle-Str. 2, 69120 Heidelberg, Germany\\$^2$ INAF-Osservatorio
     Astronomico di Bologna, Via Ranzani 1, 40127 Bologna, Italy\\$^3$
     Dipartimento di Astronomia, Universit\`a di Bologna,
     Via Ranzani 1, 40127 Bologna, Italy\\$^4$
     INFN, Sezione di Bologna, viale Berti Pichat 6/2, I-40127 Bologna, Italy}

\date{\emph{Astronomy \& Astrophysics, submitted}}

\abstract{We use simulated merger trees of galaxy-cluster halos to study the
  effect of the halo concentration distribution on strong lensing and X-ray
  emission. Its log-normal shape typically found in simulations favors
  outliers with high concentration. Since, at fixed mass, more concentrated
  halos tend to be more efficient lenses, the scatter in the concentration
  increases the strong-lensing optical depth by $\lesssim50\%$. Within cluster
  samples, mass and concentration have counteracting effects on strong lensing
  and X-ray emission because the concentration decreases for increasing
  mass. Selecting clusters by concentration thus has no effect on the lensing
  cross section. The most efficiently lensing and hottest clusters are
  typically the \textit{least} concentrated in samples with a broad mass
  range. Among cluster samples with a narrow mass range, however, the most
  strongly lensing and X-ray brightest clusters are typically $10\%$ to $25\%$
  more concentrated.}


\maketitle

\section{Introduction}

It is widely accepted now that dark-matter halos in both simulations and
reality are less concentrated, i.e.~have larger relative core sizes, the more
massive they are. This is interpreted as a consequence of hierarchical,
bottom-up structure formation. More massive halos form later, in a less dense
environment, and thus reach lower central densities. The variety of their
individual formation histories gives rise to a concentration distribution
that simulations show to be approximately log-normal with a standard
deviation of $\approx0.2$. 

What effects does this fairly broad concentration distribution have on
observable properties of galaxy clusters, most notably their strong
gravitational lensing cross sections and their X-ray temperatures and
luminosities? The log-normal distribution is substantially skewed and allows
larger positive than negative deviations from the mean. At fixed halo mass,
this should lead to outliers with higher temperature, higher X-ray luminosity,
and larger strong-lensing cross sections than expected for the nominal
concentration value. 

How are such expectations to be extrapolated to cluster samples? Above a given
mass limit, halos with lower mass and generally higher concentration are much
more abundant than more massive and typically less concentrated halos. Mass
and concentration have counter-acting effects on most observables. For
example, at fixed concentration, more massive halos are more efficient
lenses as well as hotter and more luminous X-ray emitters. However, since the
concentration is decreasing with increasing mass, these effects are at least
partially reduced. 

Here, we study the effect of the concentration distribution on several cluster
properties. We use simulated merger trees of cluster-sized, dark-matter halos,
for which concentrations are
randomly drawn from a log-normal distribution. We focus on three observable
quantities, namely the strong-lensing efficiency and the X-ray temperature and
luminosity of these clusters, and model all of them with semi-analytic
algorithms taking the importance of major mergers into account.
As a matter of fact, cluster mergers boost both lensing efficiency
and X-ray emission \citep{TO04.1,RA02.1}. 

Earlier studies on the sensitivity of strong lensing to the
concentration of dark matter halos and its scatter exist. In particular,
\cite{WY01.1}, \cite{KE01.3} and \cite{KU04.1} focused on the statistics of
multiple images as a probe of the inner structure of halos, in order to put
constraints on the dark matter self-interaction cross section, on the inner
slope of the density profile and on the equation of state parameter for dark
energy, respectively. In these studies isolated and spherical cluster models
were always  considered. In \cite{OG01.1} the effects of the concentration
and inner slope of dark matter halos on arc statistics were considered, again
assuming axial symmetry for both sources and lenses. Finally,
in \cite{HE07.1}, $N$-body simulations were used to analyse the dependence of
strong lensing cross section on several cluster properties.

The paper is organised as follows. In Sect.~2 we review the properties of the
NFW density profile, the relation between mass and concentration and different
implementations thereof. In Sect.~3, we specify the construction of the
cluster sample used in our calculations and its properties. Section~4
describes our results on the relations between halo concentration,
strong-lensing cross sections and optical depths, X-ray temperature and
luminosity. Finally, we summarise our work in Sect.~5 and discuss the
conclusions. 

\section{Dark matter halos}

\subsection{Density profile}

Quiescent dark-matter halos in $N$-body simulations acquire density profiles
well approximated by the NFW \citep{NA95.3} fitting formula (see also
\citealt{DU91.1,NA96.1,NA97.1,MO98.1,PO03.1,NA04.1}), 
\begin{equation}
\rho(r) = \frac{\rho_\mathrm{s}}{(r/r_\mathrm{s})(1+r/r_\mathrm{s})^2}\,.
\end{equation}
Its two free parameters are the scale radius $r_\mathrm{s}$, where the
logarithmic profile slope reaches $-2$, changing from $-3$ outside towards
$-1$ inside, and the scale density $\rho_\mathrm{s} =
4\rho(r_\mathrm{s})$. For a dark-matter halo at redshift $z$, $r_\Delta$ is
the radius of a sphere around the halo centre enclosing a mean density of
$\Delta$ times the critical density $\rho_\mathrm{c}(z)$ of the Universe at
redshift $z$. The mass inside $r_\Delta$ is 
\begin{equation}\label{eqn:mass}
M_\Delta = \rho_\mathrm{c} \Delta \frac{4}{3} \pi r_\Delta^3\,.
\end{equation}
According to the spherical collapse model in an Einstein-de Sitter universe,
$r_\Delta$ is the halo's virial radius at all redshifts if $\Delta = 18 \pi^2
\approx 178$ \citep{PE80.1,EK96.1}. In more general cosmologies, the virial
overdensity $\Delta$ will depend on redshift and on the cosmological
parameters. Useful fitting formulae exist \citep{LA93.1,EK96.1,BR98.2}, but we
follow the common practice to define halo masses $M_{200}$ and radii $r_{200}$
here through an overdensity of $\Delta=200$. Although they are not virial
quantities, they are used because they are independent of redshift and
cosmological parameters and adequately describe regions in virial
equilibrium. 

Accordingly, we define the concentration parameter by $c =
r_{200}/r_\mathrm{s}$. In terms of $c$, the scale radius and the scale density
can be expressed as 
\begin{equation}\label{eqn:sDen}
r_\mathrm{s} = \left(\frac{3M_{200}}{800\pi \rho_\mathrm{c} c^3}\right)^{1/3}
\quad\mbox{and}\quad
\rho_\mathrm{s} = \frac{200}{3} \rho_\mathrm{c} \frac{c^3}{F(c)}
\end{equation}
respectively, where
\begin{equation}
F(c) = \ln(1+c) - \frac{c}{1+c}\,.
\end{equation}
Halo mass and concentration can thus replace the scale radius and the scale
density as the two parameters fully describing the halo density profile. 

It has been firmly established in numerical simulations and observations
\citep{WU00.1,BU07.1,CO07.1} that the halo concentration decreases with the
halo mass. This is usually explained by the fact that low-mass halos form
earlier than massive halos in the hierarchical structure-formation scenario in
a CDM universe, and the assumption that the central halo density reflects the
mean cosmic density at the formation redshift. This explains why massive
haloes are typically found to be less concentrated than low-mass halos. The
average relation between mass and concentration allows us to characterise
halos by a single parameter, usually taken to be the virial mass $M_{200}$. 

\subsection{Concentration}

Three different algorithms were proposed in the past to relate the
concentration to the virial mass of a dark matter halo. 

The first, by \cite{NA97.1}, defines the formation redshift $z_\mathrm{c}$
of a dark-matter halo of mass $M_{200}$ collapsed at redshift $z$ as the
redshift when half of the final mass was first contained in progenitors more
massive than some fraction $f$ of $M_{200}$. 

Based on the extended Press-Schechter formalism \citep{PR74.1,BO91.1,LA93.1},
$z_\mathrm{c}$ can then be evaluated as a function of $f$, $z$ and the final
mass $M_{200}$. In line with hierarchical structure formation, NFW assumed the
scale density, which depends only on $c$ once the cosmology is fixed, to be
directly proportional to the mean matter density of the universe at
$z_\mathrm{c}$, with a proportionality constant $C$. They showed that the
$c$-$M$ relation found in a set of numerically simulated, relaxed dark matter
halos at $z=0$ is well reproduced if $f \approx 0.01$ and $C \approx 3 \times
10^3$. This holds for several different cosmological models and initial
density-fluctuation power spectra. 

\cite{BU01.1} confirmed that this algorithm works well for $z=0$, but predicts
too high halo concentrations at higher redshiftsw. They require that the
typical halo mass $M_*(z_\mathrm{c})$ at the halo-formation redshift
$z_\mathrm{c}$ be 
a fixed fraction $f$ of the final halo mass $M_{200}$. They also relate the
scale density of the halo to the critical density at the formation redshift,
but use a different definition for the scale density. The concentration found
in this way scales with redshift as $c \propto (1+z)^{-1}$, in contrast to the
much shallower redshift dependence in the NFW algorithm. 

Finally, \cite{EK01.1} proposed an alternative explanation for the $c$-$M$
relation, 
using a single parameter instead of the two parameters $C$ and $f$ and
avoiding problems of the algorithm by \cite{BU01.1} with the truncated power
spectra of Warm Dark Matter cosmogonies.They define the halo-formation
redshift $z_\mathrm{c}$ implicitly by 
\begin{equation}
D_+(z_\mathrm{c}) \sigma(M_\mathrm{s}) \left[
  -\frac{d\ln(\sigma)}{d\ln M}(M_\mathrm{s}) \right] = \frac{1}{C}\,,
\end{equation} 
where $M_\mathrm{s}$ is the mass contained within $2.17 r_\mathrm{s}$, the
radius of maximum circular velocity for the NFW density profile, $\sigma(M)$
is the standard deviation of density fluctuations on the mass scale $M$, and
$D_+(z)$ is the linear growth factor. They then equate the scale density as
defined by \cite{BU01.1} to the spherical collapse top-hat density at the
formation redshift. 

The $c$-$M$ relation by \cite{EK01.1} is probably the most general and
physically best motivated. It makes use of a single fit parameter and turned
out to reproduce halo concentrations in a variety of cosmologies, including
those with dynamical dark energy \citep{DO03.2}. It reproduces the results of
the algorithm by \cite{BU01.1} for galaxy-sized objects, but reveals
significant differences on cluster scales, as we shall see later on. 

At fixed halo mass and formation redshift, the concentration parameters of
numerically simulated dark-matter halos are log-normally distributed around
the median value $c_0$ reproduced by the algorithms described above, 
\begin{equation}\label{eqn:dist}
p(c)\d c = \frac{1}{\sigma_c \sqrt{2\pi}}\exp\left[ -\frac{(\ln c -\ln
    c_0)^2}{2\sigma_c^2}  \right]\d\ln c\,,
\end{equation}
with a standard deviation of $\sigma_c \approx 0.2$
\citep{JI00.1,BU01.1,DO03.2}. 

The log-normal distribution (\ref{eqn:dist}) is skewed towards high
concentrations. Its maximum occurs at $c_\mathrm{m} = c_0 \exp \left(
-\sigma_c^2 \right) < c_0$, and the probabilities for $c<c_0$ and $c\ge c_0$
are equal. The mean concentration is \citep{CO91.2}
\begin{equation}
\mu_1 = c_0 \exp \left( \sigma_c^2/2 \right)\,,
\end{equation}
its variance is
\begin{equation}
\mu_2 = \mu_1 \left[ \exp\left( \sigma_c^2  \right) - 1 \right]\,,
\end{equation}
and the skewness is
\begin{equation}
\mu_3 = \frac{1}{\mu_1^3} \frac{\exp \left(3\sigma_c^2 \right) -
3\exp \left(\sigma_c^2 \right)+2}
{\left[\exp \left(\sigma_c^2 \right) -1 \right]^3}\,.
\end{equation}
Setting $\sigma_c=0.2$, we find $\mu_3 \simeq 70/c_0^3 > 0$, showing that the
distribution (\ref{eqn:dist}) is substantially skewed towards high $c$. Thus
the probability of finding concentrations $c\gg c_0$ is considerably larger
than for $c\ll c_0$. This is also seen when computing the ratio of the absolute
deviations $|c-c_0|$ for $c>c_0$ and $c<c_0$, which is 
\begin{equation}
\frac{\langle |c-c_0| \rangle_+}{\langle |c-c_0| \rangle_-} =
  \frac{\mathrm{erf} 
  \left( \sigma_c/\sqrt{2} \right) + \left[ 1 - \exp \left( -\sigma_c^2/2
    \right) \right]}{\mathrm{erf}
  \left( \sigma_c/\sqrt{2} \right) - \left[ 1 - \exp \left( -\sigma_c^2/2
    \right) \right]}\,,
\end{equation} 
with the error function $\mbox{erf}(x)$. For $\sigma_c=0.2$, this ratio
becomes $\approx 1.28$, indicating that the absolute deviation for $c > c_0$
is on average $\approx 30\%$ larger than for $c < c_0$.
We shall return later to this issue to explain some of our lensing statistics
results. 

\section{Cluster population}

We model the galaxy-cluster population using one of the merger-tree sets
produced for the earlier study by \cite{FE07.1}. Extended Press-Schechter
theory was used to reproduce the formation history of $\mathcal{N} = 500$
dark-matter halos in four different dark-energy cosmologies. Here, we only use
the merger tree constructed for the concordance $\Lambda$CDM model, whose
parameters were set to $\Omega_{\mathrm{m},0} = 0.3$, $\Omega_{\Lambda,0} =
0.7$, $h = 0.65$ and $\sigma_8 = 0.84$. At redshift zero, the halos are drawn
\emph{uniformly} from the mass interval between $10^{14}$ and $2.5 \times
10^{15} M_\odot h^{-1}$ to achieve a good coverage of the mass range relevant
for strong lensing. For details on the Monte-Carlo generation of merger trees
and their applications, we refer the reader to
\cite{SO99.1,RA02.1,CA05.1,FE07.1,FE07.2}. 

Each dark-matter halo in the sample is evolved in a number of discrete
redshift steps starting from the present time up to a source redshift
$z_\mathrm{s}$ randomly drawn from the observed distribution of faint blue
galaxies parameterised by \cite{SM95.1} (see also \citealt{BA01.1}). This
distribution peaks at $z_\mathrm{s} \simeq 1.2$, rendering the region around
$z_\mathrm{l} \simeq 0.3-0.5$ the most geometrically efficient for
gravitational lensing. 

At each discrete redshift step between redshifts zero and $z_\mathrm{s}$, the
merger tree of an individual halo contains the halo mass itself and a randomly
assigned mass increment compared to the previous redshift step. This
quantifies the magnitude of the merger or smooth accretion process the halo is
undergoing within the respective time interval. 

As in \cite{FE07.1}, we twice compute the strong-lensing efficiency of each
dark-matter halo at each redshift step, first assuming that the halo can be
characterised by an unperturbed NFW density profile with elliptical
isopotential contours (we choose $e=0.3$ for the ellipticity, in agreement
with \citealt{ME03.1}), and a second time including the merger process
experienced by the halo. 

Given the mass and the redshift of a halo in the sample, we use the algorithm
by \cite{EK01.1} to compute the nominal concentration $c_0(M,z)$. Again, we
distinguish two cases in the strong-lensing analysis, assigning either the
nominal concentration $c_0$ to the halo or a value drawn randomly from the
log-normal distribution (\ref{eqn:dist}) with a standard deviation $\sigma_c =
0.2$ about $c_0$. 

We thus carry out four strong-lensing analyses for all halos in our
$\mathcal{N} = 500$ merger trees, ignoring or including the effects of merger
events and the scatter of the concentration about its nominal value set by the
$c$-$M$ relation. 

Note that this Monte-Carlo generation of merger trees should be considered as
a random experiment, representative of the evolution history of the entire
cluster population. In line with this view, we draw a new value of the
concentration at each new redshift step for each dark-matter halo. 

The lensing efficiency for a single halo is quantified by the cross section
$\sigma_\mathrm{d}$, that is the area of the domain on the source sphere in
which a source has to lie in order to produce at least one gravitational arc
with a length-to-width ratio $\ge d$. We calculate the cross sections using
the semi-analytic algorithm described in \cite{FE06.1}. It allows to
rapidly compute strong-lensing efficiencies for realistic source
distributions, and yields results that are in good agreement with those of
costly, fully-numerical ray-tracing simulations. We refer the reader to the
quoted paper for details. 

Having computed all cross sections for each of the four alternative
assumptions on the internal structure and mergers experienced by the halos, we
quantify the global lensing efficiency of the cluster population using the
optical depth per unit redshift, 
\begin{equation}\label{eqn:tau}
t_\mathrm{d}(z) =
\frac{\d\bar{\tau}_\mathrm{\d}(z)}{\d z} = \sum_{i=1}^{\mathcal{N}-1}
\frac{\sigma_\mathrm{d}(M_i,z,z_{\mathrm{s},i})}{4\pi D_{\mathrm{s},i}^2}
\int_{M_i}^{M_{i+1}} \frac{\d^2N(M,z)}{\d M\d z}\d M\,,
\end{equation}
where $D_{\mathrm{s},i}$ is the angular diameter distance to the source sphere
from the $i$-th dark matter halo in the sample, while $\d^2N(M,z)/\d M\d z$ is
the number of cosmic objects contained in the unit mass around $M$ and in the
unit redshift around $z$. The integral in (\ref{eqn:tau}) over the lens
redshift gives the total average optical depth, which is proportional to the
total number of arcs with length-to-width ratio larger than $d$ predicted to
be produced on the full sky. 

\begin{figure}[t!]
  \includegraphics[width=\hsize]{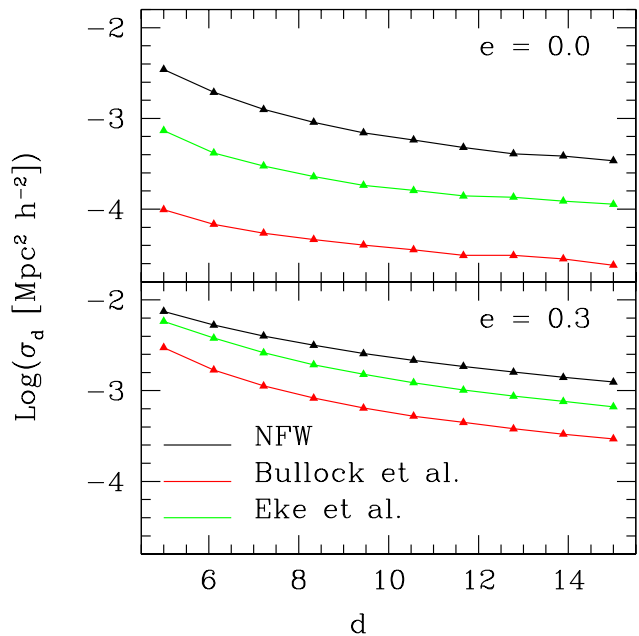}
\caption{The cross section for arcs with length-to-width ratio $\ge d$ is
  shown as a function of $d$. The mass of the lensing halo is $2 \times
  10^{15} M_\odot h^{-1}$, the lens redshift is $z_\mathrm{l} = 0.3$ and the
  source redshift is $z_\mathrm{s} = 1$. Results for three different
  prescriptions for $c$-$M$ relation are presented as labeled in the plot. The
  two panels show results for axially-symmetric (top) and elliptical (bottom)
  lenses with an isopotential ellipticity of $e=0.3$.}
\label{fig:cType}
\end{figure}
The optical depth per unit redshift is simply a sum of the cross sections of
each individual halo, weighted by the abundance of such halos at the
corresponding redshift. Weighting by the mass function causes this sum to be
dominated by the halos with the lowest masses that are still capable of
producing a non-vanishing arc cross section. Introducing the scatter into the
mass-concentration relation can lift low-mass halos above or push them below
the strong-lensing threshold. However, the skewness of the concentration
distribution makes it more likely that low-mass halos are lifted above the
threshold than the reverse. Thus, it is plausible that the log-normal
concentration distribution may have a potentially significant effect on the
strong-lensing optical depth. 

\section{Results}

\subsection{Different concentration prescriptions}

Before we continue, it is interesting to assess how the strong-lensing cross
sections differ for the different $c$-$M$ relation algorithms outlined in
Sect.~2. At the same mass and redshift, higher concentrations should push the
critical curves of a lensing halo outwards, thus increasing its strong-lensing
cross section. 

Results are shown in Fig.~\ref{fig:cType}, where we plot the cross section for
gravitational arcs with length-to-width ratios $\ge d$ as a function of $d$,
using the three algorithms for the $c$-$M$ relation. We also show the
difference between axially-symmetric and elliptical lenses. 

Evidently, the impact of different concentrations is much reduced for
elliptical compared to circular lenses. For example, if we focus on $d=10$, we
note that the cross sections differ by a factor of $\approx 4$ for elliptical
lenses. For axially-symmetric lenses, this factor grows up to $\approx 20$.
This
is owed to the fact that halo ellipticity largely increases the strong-lensing
cross section \citep{ME03.1,OG03.1,ME07.1}, causing the lensing efficiency to
be less sensitive to the internal structure of the lens. 

Next, we see that the original NFW prescription for the $c$-$M$ relation
yields the largest cross sections for all values of $d$. As explained in
Sect.~2, this is because the NFW prescription performs well at redshift zero,
but overpredicts concentrations at higher redshift. At $z=0.3$, where we
placed the lens, the concentration is thus substantially overestimated,
resulting in a very large cross section. 

Concentrations computed using \cite{BU01.1} and \cite{EK01.1} algorithms agree
on galactic scales, but differ on cluster scales. Although results obtained
with them both fall below the NFW result, they produce quite different cross
sections for all $d$. In particular, the \cite{EK01.1} algorithm yields
results falling in between those obtained with the NFW and \cite{BU01.1}
prescriptions, respectively. 

\begin{figure*}[ht!]
  \includegraphics[width=0.45\hsize]{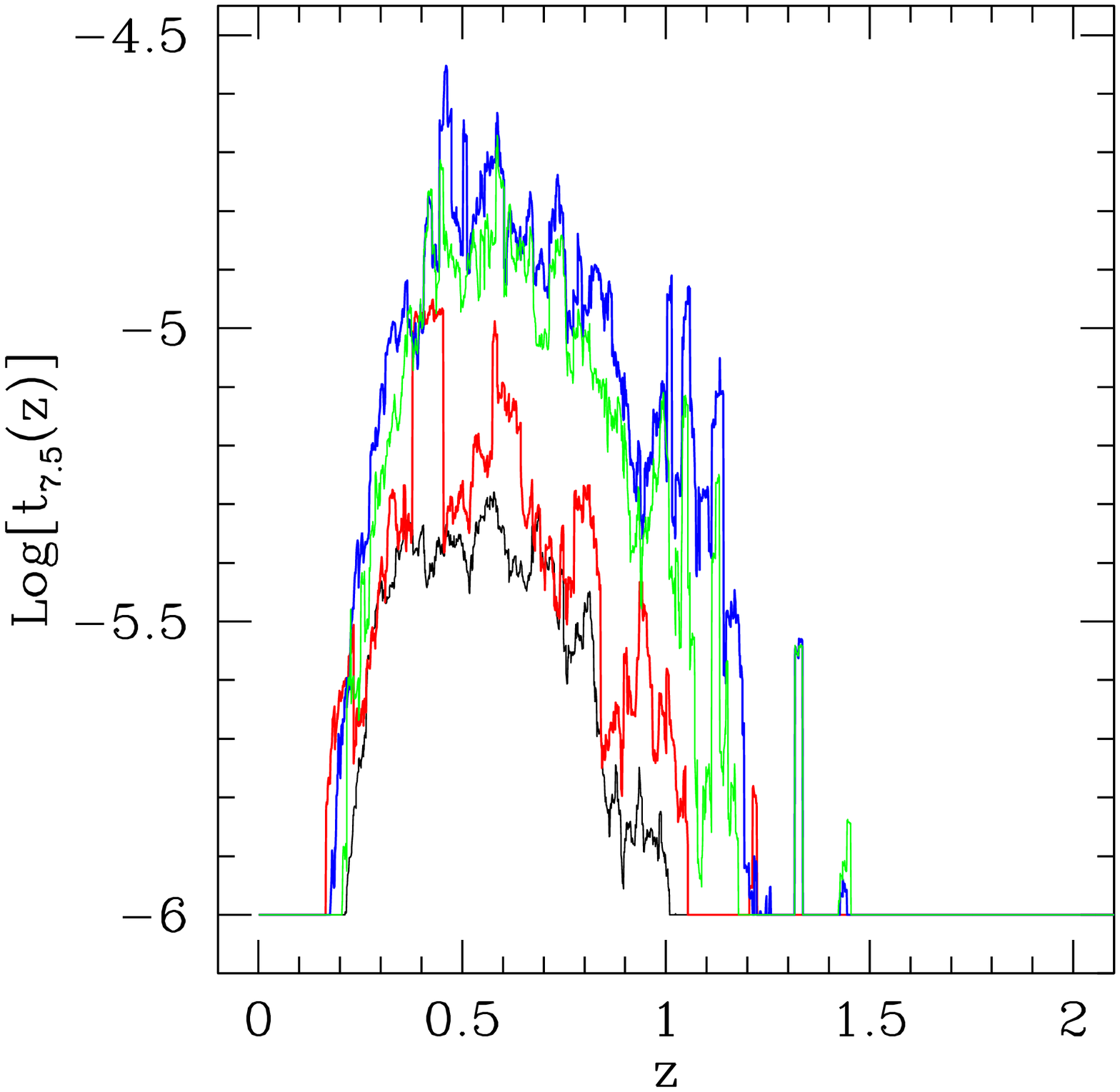}\hfill
  \includegraphics[width=0.45\hsize]{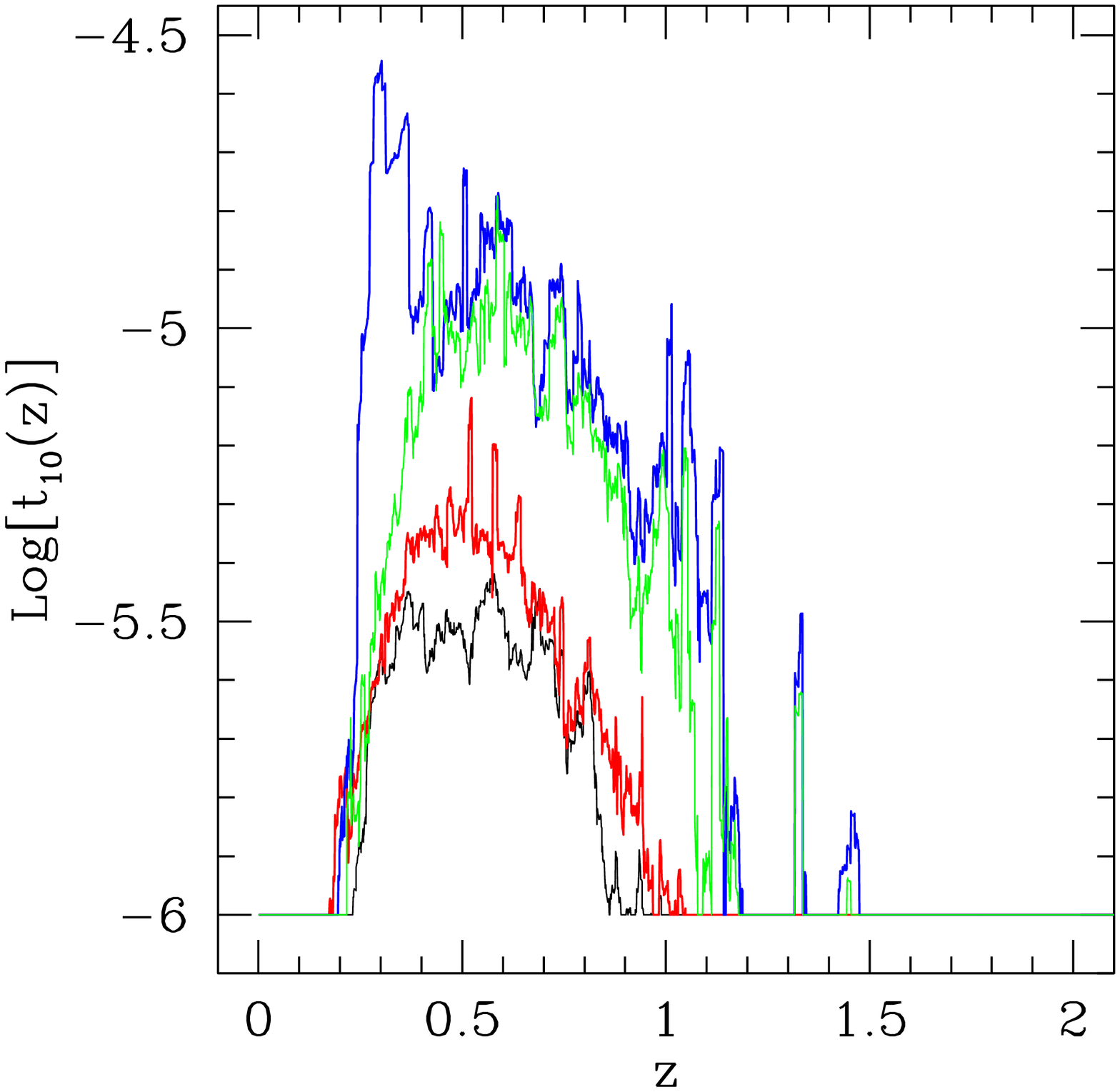}
\caption{\emph{Left panel}. Optical depth per unit redshift for arcs with
  length-to-width ratio $d\ge7.5$ as a function of the lens redshift. The thin
  black and green lines show the results obtained ignoring and accounting for
  cluster mergers, respectively, both using the nominal $c$-$M$ relation. The
  heavy red and blue curves include the scatter in the $c$-$M$
  relation. \emph{Right panel.} Similar to the left panel, but for arc
  length-to-width ratios $d\ge10$, and using a different random-number seed.}
\label{fig:tau}
\end{figure*}

This illustrates that the choice of the $c$-$M$ relation is very important in
analytic and semi-analytic models of galaxy-cluster lensing since different
concentrations can have a large effect on the strong-lensing properties. The
factors exceeding one order of magnitude between different prescriptions shown
in Fig.~\ref{fig:cType} for axially symmetric lenses is particularly striking
in this regard. 

We compared strong-lensing cross sections for several dark-matter halos
extracted from a high-resolution numerical simulation with those of analytic
lens models with NFW density profile with the same mass and redshift, an
isopotential ellipticity of 0.3 and with each of the three different
algorithms for the $c$-$M$ relation. We generally find the best agreement of
the strong-lensing efficiencies for concentrations computed with the algorithm
by \cite{EK01.1}. This further supports the plausibility of this algorithm for
the $c$-$M$ relation. From now on, we assign fiducial concentrations by means
of the \cite{EK01.1} algorithm for the $c$-$M$-relation. 

\subsection{Scatter in the concentration}

We now proceed as anticipated in Sect.~3, performing four different
strong-lensing analyses for our dark-matter halo population. 

We show in Fig.~\ref{fig:tau} the optical depth per unit redshift as a
function of lens redshift as defined in Eq.~(\ref{eqn:tau}), for arcs with
length-to-width ratios $d\ge7.5$ and $d\ge10$ respectively. Results are shown
both including and ignoring the effect of cluster mergers, and both assuming
the ideal $c$-$M$ relation and introducing a concentration scatter consistent
with the log-normal distribution of Eq.~(\ref{eqn:dist}). 

For the two cases $d\ge7.5$ and $d\ge10$, we used two different seeds for
drawing random concentrations from the distribution in order to gain insight
into the effect of limited statistics. 

We first note the general trend that the introduction of the scatter in the
$c$-$M$ relation systematically increases the optical depth, and this is true
irrespective of whether halo mergers are taken into account or ignored. This
is a consequence of the skewness of the concentration distribution;
cf.~Sect.~2. Since concentrations much larger than the fiducial value are more
probable than much lower concentrations, it is more likely for the
concentration scatter to increase the strong-lensing cross section rather than
the reverse. In other words, halo concentrations become larger on average after
introducing the scatter, thus producing a larger optical depth per unit
redshift. 

In closer detail, we note several local maxima of the differential optical
depths per unit redshift obtained after introducing a scatter in the $c$-$M$
relation. These are caused by individual dark-matter halos with relatively low
mass that, due to the random assignment of concentrations, reach a
particularly high concentration and thus a large cross section. Because of
their low mass, they have a large relative abundance, thus they dominate the
sum in the optical depth per unit redshift, Eq.~(\ref{eqn:tau}), and cause the
peaks. 

The position, width and amplitude of these peaks change of course if the seed
for the random-number generation is changed. However, even though the
\textit{local} increase in the differential optical depth can be quite
significant, the increase in the \textit{total} optical depth, i.e.~the
integral under the curves in Fig.~\ref{fig:tau}, is limited to
$\approx 40-50\%$, both including or ignoring halo mergers. 

To study this in more detail, we concentrate on $d\ge10$ and the more
realistic case when mergers are taken into account. We further select a halo
subsample with redshifts between $z_1=0.28$ and $z_2=0.32$, centred on
$z=0.3$. Since our original cluster sample was randomly drawn from a uniform
mass distribution at $z=0$ and then evolved backwards in time to construct
merger trees, each dark-matter halo of mass $M_{200}$ at redshift $z$ needs to
be statistically weighted by the abundance of such halos according to the mass
function for the cosmological model at hand. We note that appropriate weights
are included in the optical-depth calculations, see Eq.~(\ref{eqn:tau}). 

\begin{figure}[t!]
  \includegraphics[width=\hsize]{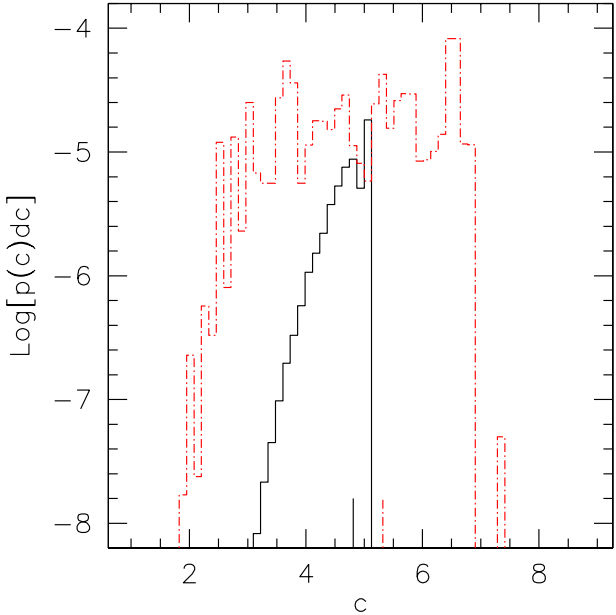}
\caption{Unnormalised distribution of the concentrations for all the halos in
  our sample with redshifts between $z_1=0.28$ and $z_2=0.32$. The black solid
  histogram shows the result obtained adopting the fiducial $c$-$M$ relation
  of \cite{EK01.1}. The log-normal concentration scatter is taken into account
  for the red-dashed histogram. The vertical dashes indicate the median
  concentration in both cases.} 
\label{fig:cDist}
\end{figure}

\begin{figure}[t!]
  \includegraphics[width=\hsize]{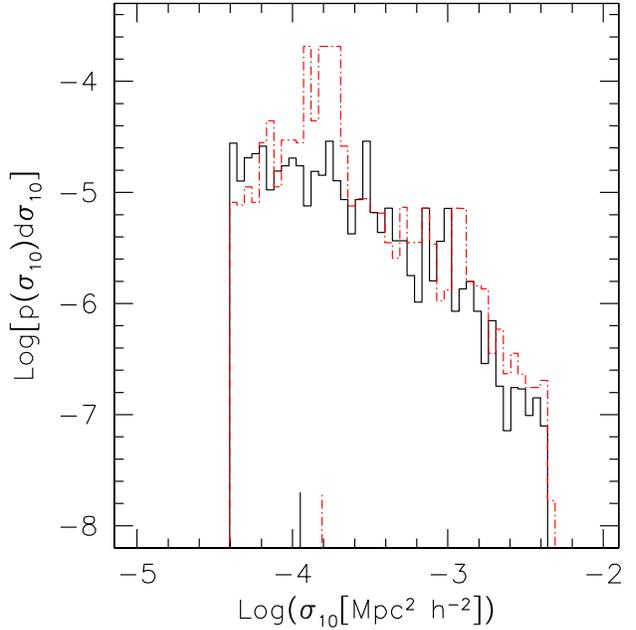}
\caption{Unnormalised distribution of the cross sections for gravitational
  arcs with length- to-width ratios $d\ge10$ for all halos in our subsample
  with redshifts between $z_1=0.28$ and $z_2=0.32$. As in
  Fig.~\ref{fig:cDist}, the black solid  and red dashed histograms show
  results ignoring the concentration scatter and accounting for it,
  respectively. Dashed vertical lines mark the median cross sections for both
  cases.} 
\label{fig:sDist}
\end{figure}

Figures~\ref{fig:cDist} and \ref{fig:sDist} show the distributions of
concentrations and strong-lensing cross sections in the halo subsample. In
both figures, we contrast results obtained ignoring the concentration scatter
(solid black curves) and taking it into account (red dashed curves). Note that
all distributions shown are unnormalised. 

Without scatter, the concentration distribution is very peaked, but it
flattens and widens when the scatter is taken into account, as one would
expect. Note also that both concentration distributions drop very sharply at
high concentrations. This reflects the mass cutoff in our halo sample, since
high concentrations correspond to low masses. 

The cross-section distributions behave similarly. However, in this case the
sudden cut-off at low cross sections is due to the strong-lensing
threshold. For producing large arcs, a halo's caustics need to be sufficiently
larger than the available sources. Below this threshold, the strong-lensing
cross sections sharply drop to zero. See also \cite{FE06.1} for more
discussion of this issue and its implementation. 

Finally, the systematic increase of the differential optical depth shown if
Fig.~\ref{fig:tau} can be further
understood as the contribution of two factors. First, we note that the median
concentration (and hence also the median strong-lensing cross section) is
larger when the concentrations scatter about the mean $c$-$M$
relation. Second, the significant peaks in Fig.~\ref{fig:sDist} (note the
logarithmic scale!) appearing in the cross-section distribution at relatively
low cross sections are produced by rather low-mass halos that dominate the sum
in the optical depth per unit redshift because of their large statistical
weight. 

\subsection{Lensing concentration bias}

Another interesting issue that we are able to explore with our halo sample
regards the strong-lensing cross sections expected for concentrated halos, and
conversely the concentrations expected in efficient strong-lensing halos. 

This will allow us to better understand the relative effect of mass and
concentration on the amplitude of the strong lensing cross section, and to
quantify the bias expected to be found in dark-halo concentration measurements
of strongly-lensing clusters. We can then compare such results to those
obtained by \cite{HE07.1}, who carried out among other things a similar
analysis on a large set of numerically simulated dark-matter halos. 

\begin{figure}[t!]
  \includegraphics[width=\hsize]{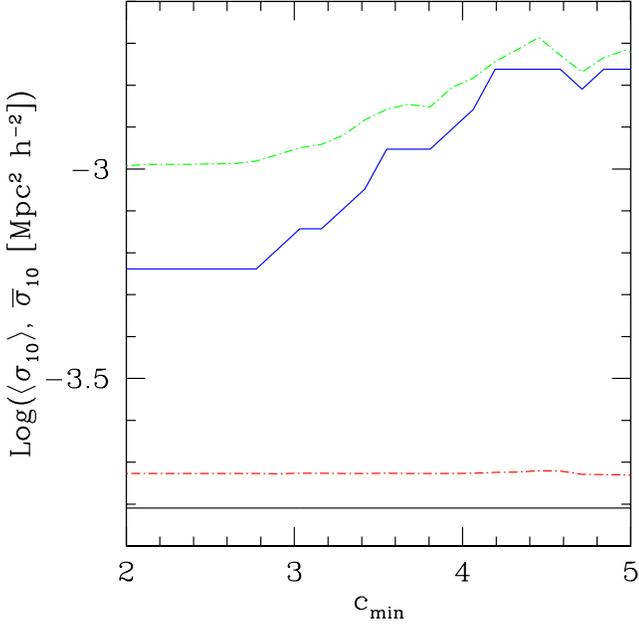}
\caption{The black and blue solid curves show the median, and the red and
  green dashed curves the mean cross section for arcs with length-to-width
  ratio $d\ge10$. Only halos with concentrations above the threshold on the
  abscissa are included. The bottom pair of lines shows the result without any
  mass selection, while only halos more massive than $7.5 \times 10^{14}
  M_\odot h^{-1}$ are included in the top pair.}
\label{fig:avSigma}
\end{figure}

Figure~\ref{fig:avSigma} shows the median $\overline{\sigma}_{10}$ and the mean
$\langle \sigma_{10} \rangle$ cross sections of the halo subsample, restricted
to those halos with a concentration exceeding the threshold on the
abscissa. Results are shown both for all halos irrespective of their mass, and
only for halos with masses $\ge7.5 \times 10^{14} M_\odot h^{-1}$. 

Without mass selection, the curves are flat within the range of concentrations
shown. Remarkably, this indicates that low-mass halos with their typically
high concentrations have similar mean or median cross sections as high-mass
halos and therefore contribute most of the strong-lensing optical depth in the
halo subsample because of their high abundance. 

This result may seem at odds with the expectation that the lensing efficiency
should increase with increasing halo concentration, as illustrated in
Fig.~\ref{fig:cType} when we discussed the effect of different algorithms
implementing the $c$-$M$ relation. However, note that Fig.~\ref{fig:cType}
shows results for a single halo mass. If we select only the most massive
halos, we find an increase of the mean and median cross sections with the
concentration threshold. Thus, once the mass dependence is effectively
suppressed in this way, the concentration dependence of the strong-lensing
efficiency can emerge. In other words, although the average strong-lensing
cross sections do indeed increase with the halo concentration, this effect is
almost precisely cancelled if halos of all masses in a broad mass range are
considered. 

According to Fig.~\ref{fig:avSigma}, the median and mean cross sections of
massive halos can increase by a factor of $\approx 2.5$ as the concentration
increases from $2$ to $5$. 

\begin{figure}[t!]
  \includegraphics[width=\hsize]{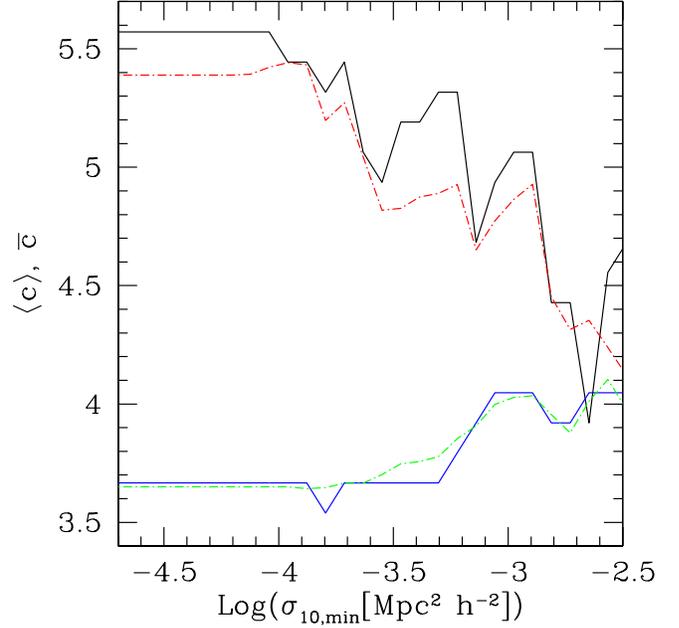}
\caption{The black and blue solid curves show the median, the red and green
  dashed curves the mean concentration. Only halos with strong-lensing cross
  sections above the threshold on the abscissa are taken into account. The top
  pair of curves shows the result obtained without mass selection, while only
  halos more massive than $7.5\times 10^{14} M_\odot h^{-1}$ contribute to the
  bottom pair of curves.} 
\label{fig:avC}
\end{figure}

Figure~\ref{fig:avC} shows the mean $\langle c\rangle$ and median $\bar{c}$
concentration of halos with strong-lensing cross sections above the threshold
on the abscissa. Again, we compare the complete halo subsample with massive
halos above a mass limit of $7.5\times 10^{14} M_\odot h^{-1}$. We note that
(i) if we impose no mass threshold, the concentration for strongly lensing
halos is always smaller on average compared to the entire population, and (ii)
if we allow only massive halos, the mean and median concentrations increase
with the lensing cross section. 

Specifically, the mean and median concentrations of massive halos shown in
Fig.~\ref{fig:avC} increase by $\approx 12\%$ across the range of
cross-section thresholds shown. If we further raise the mass threshold, the
increase rises to $\approx 25\%$. 

Without any mass selection, the highest cross sections are produced by the
most massive objects, that are on average less concentrated than the low-mass
halos. If we restrict the analysis to massive halos, we remove part of the
mass dependence of the strong-lensing efficiency and find that the
concentrations found in strongly lensing clusters are slightly biased
high. Narrowing the mass interval, the effect of the concentration is less
diluted by the mass dependence, thus increasing the bias. This result agrees
with the corresponding result of \cite{HE07.1} and will be discussed later
on. 

\subsection{X-ray concentration bias}

It is now interesting to ask whether comparable concentration biases are
expected in X-ray selected cluster samples. At fixed mass, a more concentrated
halo creates a deeper potential well and thus causes the intracluster gas to
become hotter in thermal and hydrostatic equilibrium. The gas density will
also increase, thus raising the X-ray luminosity. 

To address this question, we first require a relation between the X-ray
observables and mass, the redshift and the concentration of the host
dark-matter halo. We achieve this following \cite{EK98.1} who derived an
extension to the usual cluster scaling relations
\citep{WH78.1,WH82.1,KA86.1}. First of all, the circular velocity profile for
a dark-matter halo with an NFW density profile is \citep{NA97.1} 
\begin{equation}
\left[\frac{v(r)}{v_{200}}\right]^2 = \frac{r_{200}}{r}
\frac{F(cr/r_{200})}{F(c)}\,,
\end{equation}
where $v_{200}$ is the circular velocity at $r_{200}$, that is $v^2_{200} =
GM_{200}/r_{200}$. This distribution peaks at $r \approx 2 r_{200}/c$,
corresponding to 
\begin{equation}
v_\mathrm{m}^2 \approx 0.22 v^2_{200} \frac{c}{F(c)}\,.
\end{equation}
This characteristic velocity of the system measures the depth of its potential
well. If only gravity or other scale-free processes like pressure gradients or
hydrodynamical shocks dominate within the cluster, any other measure of the
potential depth, such as the temperature of the intra-cluster gas, must be
proportional to $v_\mathrm{m}^2$, that is 
\begin{equation}\label{eqn:tempPro}
T(M_{200},z,c) \propto \frac{M_{200}}{r_{200}} \frac{c}{F(c)}\,.
\end{equation} 
Now, Eq.~(\ref{eqn:mass}) implies
\begin{equation}
r_{200} = \left[ \frac{3M_{200}}{800 \pi \rho_c(z)}  \right]^{1/3}\,.
\end{equation}
Inserting this into Eq.~(\ref{eqn:tempPro}), we can write
\begin{equation}\label{eqn:temp}
T(M_{200},z,c) = \beta \left[M_{200}h(z)\right]^{2/3}\frac{c}{F(c)}\,,
\end{equation}
where $\beta$ collects now all the constant factors. Note that this relation
retains the mass and redshift dependence of the temperature of the common
scaling relation, but acquires the concentration dependence from the
dark-matter density profile. In particular, the function $c/F(c)$ is a
monotonically increasing function of the concentration if $c \gtrsim 2$, which
is almost always the case in our halo sample (cf.~the concentration
distribution in Fig.~\ref{fig:cDist}). 
It is shown that adiabatic simulations of gas in galaxy clusters
follow relatively well this type of scaling relation
\citep{EK98.1,BR98.2}. With the introduction of more complex physical
processes, like non gravitational heating and radiative cooling, the scaling
relation is instead not closely reproduced \citep{BA02.2,KA02.1}. However,
in spite of simplicity, we prefer to stick to it, leaving more complicated
models for further study.

\begin{figure}[ht!]
  \includegraphics[width=\hsize]{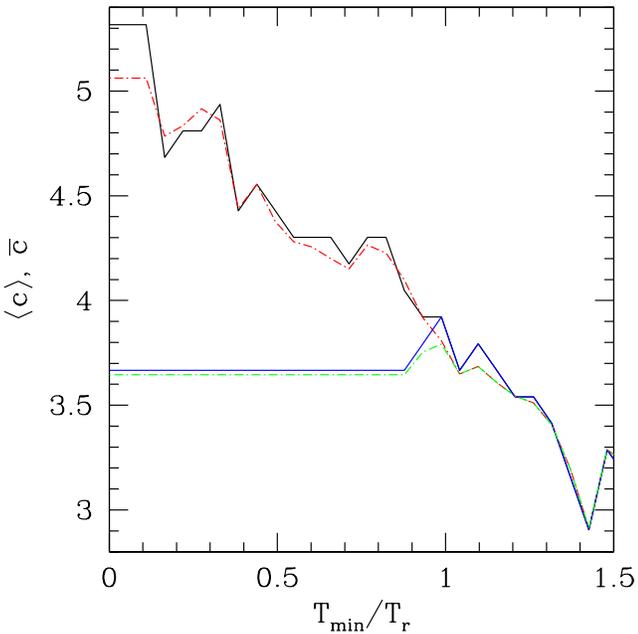}
\caption{Black and blue solid curves show the median, red and green dashed
  curves the mean concentration for the subsample of dark-matter halos between
  $z_1=0.28$ and $z_2=0.32$. Only halos with relative temperatures exceeding
  the threshold on the abscissa are included. The top pair of curves shows the
  result without mass selection, while only halos with mass larger than
  $7.5 \times 10^{14} M_\odot h^{-1}$ contribute to the bottom pair.} 
\label{fig:xRay}
\end{figure}

\begin{figure}[ht!]
  \includegraphics[width=\hsize]{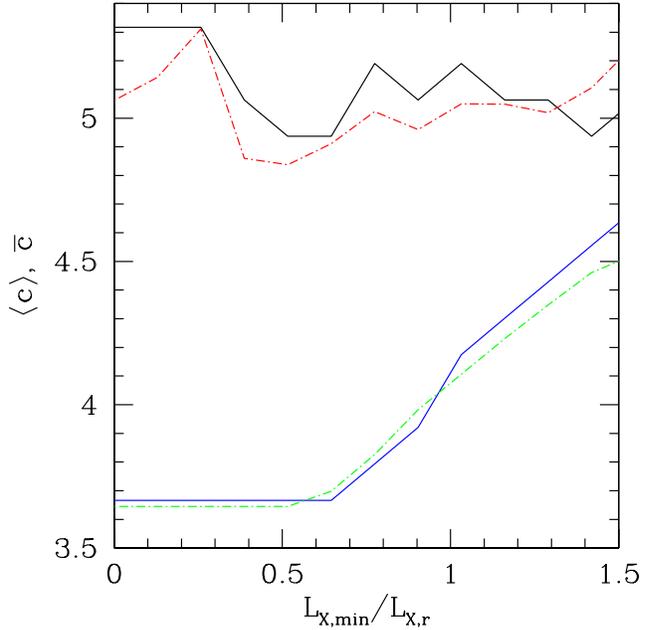}
\caption{Black and blue solid curves show the median, red and green dashed
  curves the mean concentration for the subsample of dark-matter halos between
  $z_1=0.28$ and $z_2=0.32$. Only halos with relative X-ray luminosities
  exceeding the threshold on the abscissa are taken into account. The top pair
  of curves was obtained without mass selection, while only massive halos with
  mass larger than $7.5 \times 10^{14} M_\odot h^{-1}$ contribute to the
  bottom pair.} 
\label{fig:xLum}
\end{figure}

Quantifying the bolometric X-ray luminosity of the intra-cluster gas, we start
from 
\begin{equation}\label{eqn:lum}
L_X(M_{200},z,c) = 4\pi \int_0^{+\infty} r^2 \rho_\mathrm{g}(r)^2
\frac{\Lambda(T)}
{(\mu m_\mathrm{p})^2}\d r\,,
\end{equation}
where $\Lambda(T)$ is the cooling function, depending on the relevant
radiative processes, and $\rho_\mathrm{g}(r)$ is the gas-density profile. We
assume that the gas density follows the dark matter density, $\rho_\mathrm{g}
= f_\mathrm{g} \rho$, with a constant factor constant $f_\mathrm{g}$. This is
of course not strictly true, especially in the inner region where the
dark-matter density profile is cuspy while the gas distribution forms a finite
core due to the gas pressure. However, the final result is insensitive to this
simplifying assumption. Further assuming that the intracluster gas is
isothermal, the luminosity can be written as 
\begin{equation}
L_X(M_{200},z,c) = 200
\Lambda(T) \left(\frac{f_\mathrm{g}}{3\mu m_\mathrm{p}} \right)^2
M_{200} \rho_\mathrm{c}(z) \frac{c^3}{F(c)^2}\,.
\end{equation}
If the main emission mechanism of the intra-cluster gas is thermal
bremsstrahlung, then $\Lambda(T) \propto T^{1/2}$. Hence, recalling
Eq.~(\ref{eqn:temp}) and collecting all constant factors into $\gamma$, we get 
\begin{equation}\label{eqn:lumi}
L_X(M_{200},z,c) =
\gamma M_{200}^{4/3} h(z)^{7/3} \frac{c^{7/2}}{F(c)^{5/2}}\,.
\end{equation}
The common dependence of the luminosity on the mass and the redshift of the
host dark matter halo is retained again, and an additional dependence on the
concentration appears. Note that the concentration dependence is steeper here
than for the temperature.
Note that the dependence of the bolometric X-ray luminosity on
the concentration shown in Eq.~(\ref{eqn:lumi}) differs by a factor of
$1 - (1+c)^{-3}$ from the formula given in
\cite{EK98.1}. This is because the integral in
Eq.~(\ref{eqn:lum}) extends to infinity, while it was
limited to the virial radius in \cite{EK98.1}.
This is unimportant because the missing factor is very close to
unity for all reasonable values of the concentration.

In the following, we refer the temperature and the X-ray luminosity of the gas
inside each dark-matter halo of our subsample to the temperature
$T_\mathrm{r}$ according to (\ref{eqn:temp}) and the luminosity
$L_{X,\mathrm{r}}$ according to (\ref{eqn:lumi}) of a reference halo with mass
$M_{200,\mathrm{r}} = 10^{15} M_\odot h^{-1}$ placed at redshift $z_\mathrm{r}
= 0$. It has a nominal concentration $c_\mathrm{r} = 3.74$ according to the
\cite{EK01.1} algorithm. Thus, for each halo, we only consider the relative
temperature 
\begin{equation}
\frac{T(M_{200},z,c)}{T_\mathrm{r}} =
\left[\frac{M_{200}h(z)}{M_{200,\mathrm{r}}h}\right]^{2/3} \frac{c}{F(c)}
\frac{F(c_\mathrm{r})}{c_\mathrm{r}}\,,
\end{equation} 
and the relative luminosity
\begin{equation}
\frac{L_X(M_{200},z,c)}{L_{X,\mathrm{r}}} =
\left(\frac{M_{200}}{M_{200,\mathrm{r}}}\right)^{4/3}
\left[\frac{h(z)}{h}\right]^{7/3} 
\frac{c^{7/2}}{F(c)^{5/2}} \frac{F(c_\mathrm{r})^{5/2}}{c_\mathrm{r}^{7/2}}\,.
\end{equation}
Figure~\ref{fig:xRay} shows the median and mean concentrations for dark-matter
halos with a relative gas temperature exceeding the threshold on the
abscissa. We show the results both without any mass selection and selecting
halos more massive than $7.5 \times 10^{14} M_\odot h^{-1}$. Evidently, the
mean and median halo concentrations decrease in both cases as the relative
temperature threshold increases. This illustrates that particularly hot gas
resides in the most massive halos, quite irrespective of the
concentration. Also, if we consider only the most massive objects, a plateau
appears at low temperatures because low-temperature clusters are then removed
from the sample. Thus, the gas temperature depends so weakly on the halo
concentration compared to its dependence on mass that even a narrow mass
selection does not reveal the increasing concentration-temperature relation. 

Figure~\ref{fig:xLum} shows the mean and median concentrations in halos
selected for their X-ray luminosity. If all halos in the subsample are
included, the curves are almost flat, showing that the concentrations are
typically independent of the X-ray luminosity. If only massive halos are
included, the mean and median concentrations increase such that the most
luminous X-ray clusters can be up to $\approx 25\%$ more concentrated than the
entire cluster population. 

Hence, unlike for the temperature, we here find increasing mean and median
concentrations as a function of the luminosity threshold. In summary, a
concentration bias in temperature-selected clusters is not expected, but the
most massive and X-ray luminous clusters are typically more concentrated than
the population of X-ray clusters indicating a concentration bias similar to
that found in strongly-lensing clusters. 

The different results for clusters selected by temperature or X-ray luminosity
can be understood considering the following numbers. As remarked before, the
nominal concentration of the reference cluster is $c_\mathrm{r} = 3.74$. Had
we adopted a reference mass of $2.5 \times 10^{14} M_\odot h^{-1}$, the
nominal concentration was $c_\mathrm{r} = 4.73$. These two concentrations are
1-$\sigma$ compatible with the same underlying mass, given the variance of
$\sigma_c = 0.2$ in the log-normal concentration distribution. The increase in
the gas temperature due to the higher concentration is only $\approx 5\%$,
while the X-ray luminosity increases by $\approx 45\%$. On the other hand, the
gas temperature drops by a factor of $\approx 2.5$ because of the lower halo
mass,
while the bolometric X-ray luminosity drops by a factor of $\approx 6.3$. On
the whole, the ratio between the changes in temperature due to the halo mass
and due to the concentration is $\approx 12$, while the ratio between the
changes in X-ray luminosity due to the mass and due to the concentration is
$\approx 1.9$. This shows that the effect of the concentration on the X-ray
luminosity is almost comparable to the effect of the mass, but much less
important for the temperature. 

In other words, the mass dependence of the gas temperature is overwhelmingly
stronger than its concentration dependence, cancelling any kind of
concentration bias that could appear in temperature-selected halos. Very hot
clusters are actually less concentrated (more massive) than average. On the
other hand, the stronger dependence of the luminosity on the concentration
allows to invert this trend if only massive clusters are considered. Thus,
very X-ray luminous clusters have higher mean and median concentrations
than clusters with lower luminosity but comparable mass. 

\begin{figure}[t!]
  \includegraphics[width=0.45\hsize]{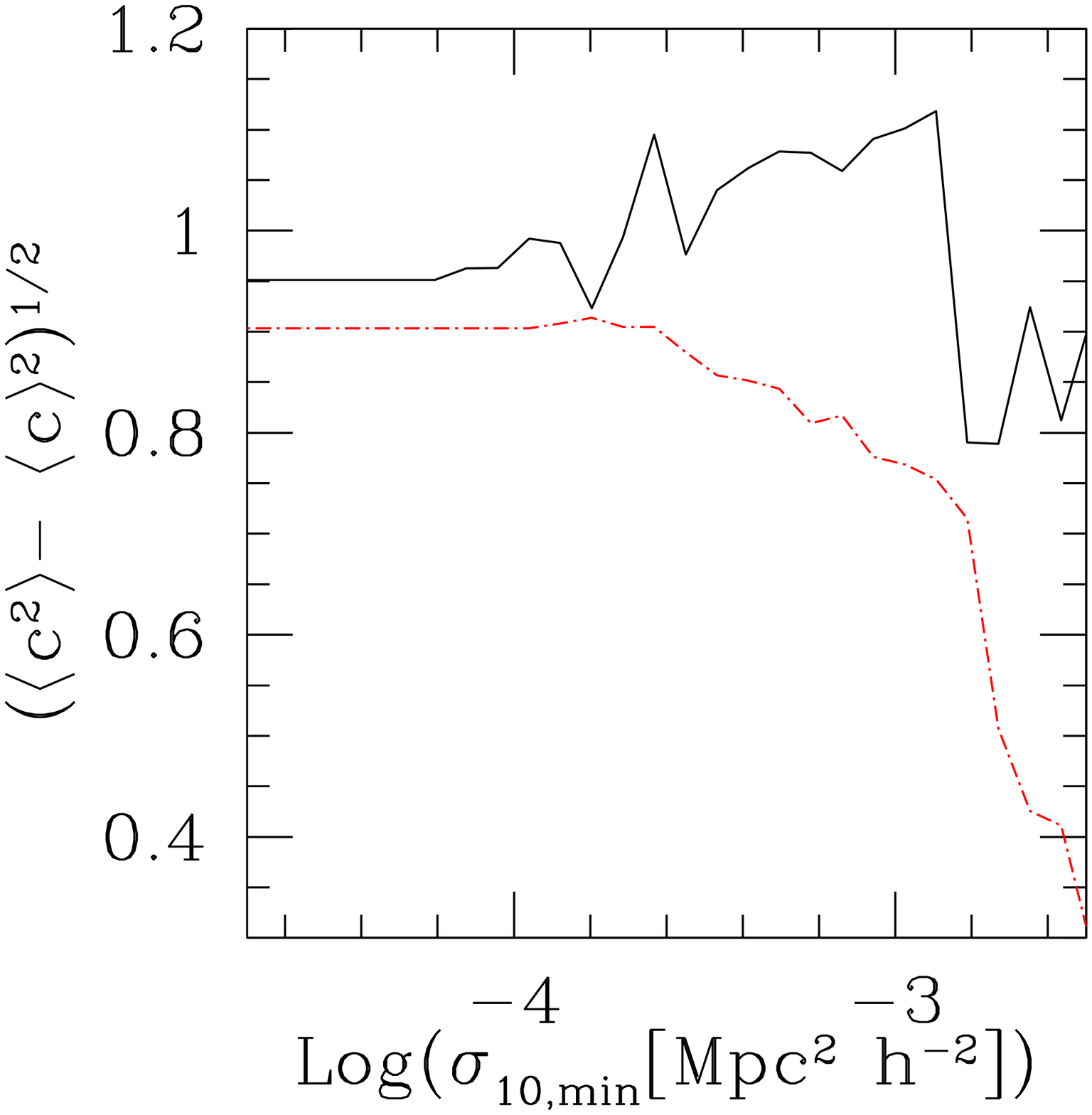}\hfill
  \includegraphics[width=0.45\hsize]{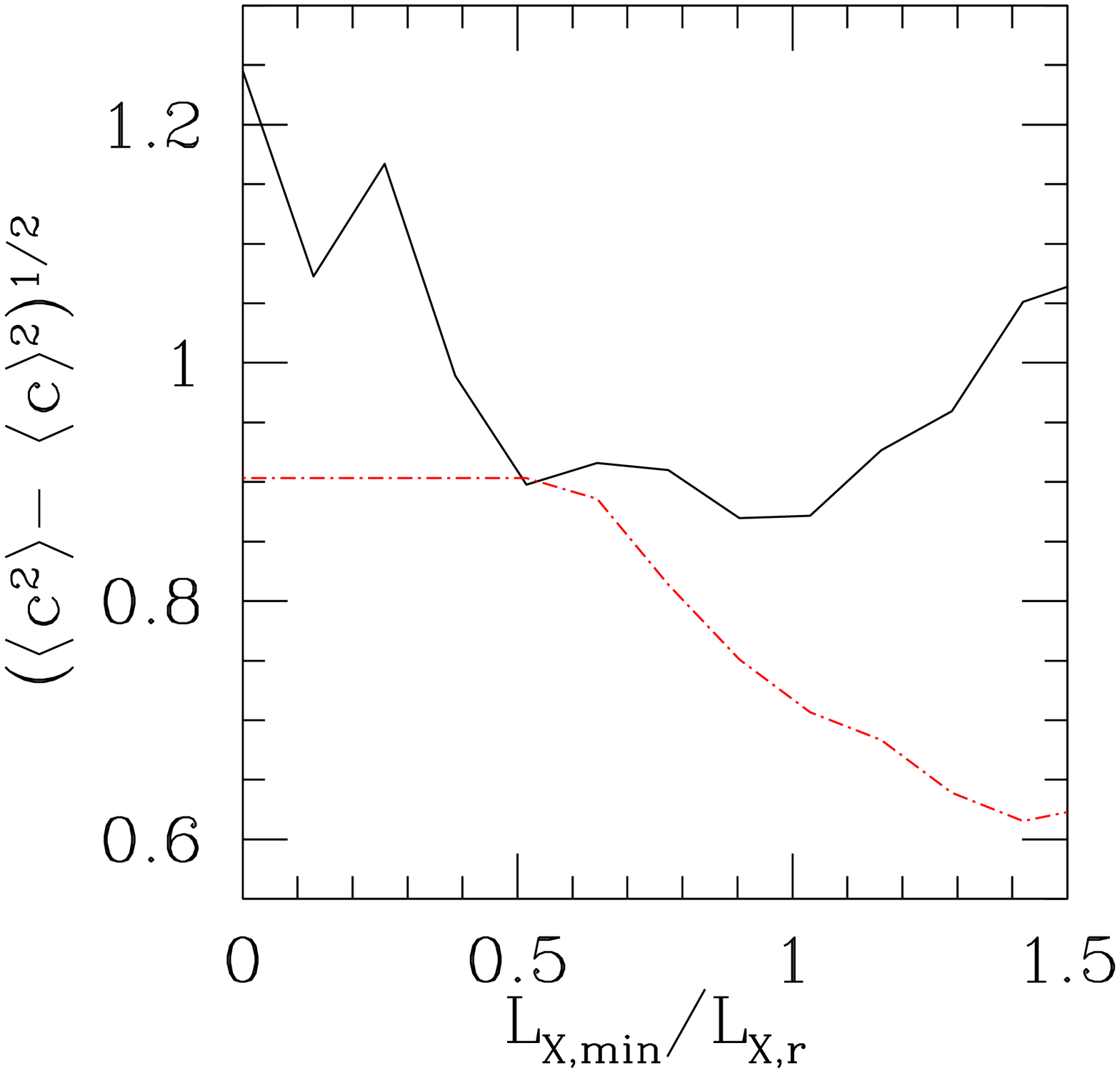}
\caption{The \textit{rms} of the concentration distribution accounting for
  halos in the subsample with strong-lensing cross sections (left panel) or
  relative X-ray luminosities (right panel) exceeding the thresholds on the
  abscissa. The solid black lines are obtained without mass selection, while
  only massive halos with mass larger than
  $7.5 \times 10^{14} M_\odot h^{-1}$ contribute to the red dashed curves.} 
\label{fig:rms} 
\end{figure}

To see which concentrations we can expect in suitably selected cluster
samples, we plot in Fig.~\ref{fig:rms} the \textit{rms} $(\langle c^2\rangle -
\langle c\rangle^2)^{1/2}$ of the concentration distribution as a function of
the cross-section and X-ray luminosity thresholds, respectively, both with and
without further mass selection. According to Figs.~\ref{fig:avC} and
\ref{fig:xLum}, the median and the mean of the distribution are quite similar,
hence the distribution itself is quite symmetric, and the \textit{rms} is a
good estimator of its width. 

Without mass selection, the \textit{rms} always remains around unity. If we
introduce mass selection, it is close to unity for the entire subsample, but
drops towards $0.4$ when only efficient strong lenses are included, and to
$0.6$ when only very X-ray luminous clusters are included. This means that the
concentration distribution tends to narrow in the latter cases. 

\subsection{Additional effects}

Finally, we explore the consequence for our results of two
additional effects not included so far. The first is the correlation
of the concentration with the triaxiality of dark-matter halos
\citep{JI02.1}. The second is the ellipticity distribution of projected
halos due to the random orientation of the three-dimensional
halos with respect to the line-of-sight \citep{OG03.1,OG05.1,CO06.1}.
The second effect affects only
the strong lensing properties of galaxy clusters, for whose lensing potential
we assumed an ellipticity of $e = 0.3$ throughout this work.
The scaling laws we used for the X-ray characteristics are insensitive
to the ellipticity of the dark-matter halo. Besides, the gas distribution
approximately follows equipotential surfaces and thus tends to be more
spherical than the dark matter distribution \citep{GA05.1}.

We assess the impact of these two effects
in the following experiment. First, we considered a dark-matter halo
with mass
$2 \times 10^{15} M_\odot h^{-1}$ and redshift $z_\mathrm{l} = 0.3$.
We computed its cross section for arcs with length-to-width
ratio $d \ge 10$, assuming sources at $z_\mathrm{s} = 1$, a lensing-potential
ellipticity $e = 0.3$ and concentration derived from the algorithm of
\cite{EK01.1}. Then, we produced 1,000 triaxial modifications of this
original halo by drawing axis ratios from the
distributions given in \cite{JI02.1}. The axis ratios allow changing the
concentration of each modified halo
according to the prescription of \cite{JI02.1}, predicting higher
concentrations for more spherical halos. Finally, each modified
halo is projected along a randomly selected line-of-sight and
the ellipticity of the projected
density is computed following \cite{OG03.1}. To each halo is then assigned
a new lensing-potential ellipticity assuming that it is half of the
ellipticity of the projected density.

As outlined in \cite{JI02.1}, the isodensity surfaces tend to be more
elongated near the core of the halo than in its outer regions.
Since the innermost part of a galaxy cluster is most relevant for
strong-lensing events, we lowered the minor-to-major and
intermediate-to-major axis ratios by 0.15 prior to the projection.
This is consistent with Fig.~3 of \cite{JI02.1}.

\begin{figure}[t!]
  \includegraphics[width=1.0\hsize]{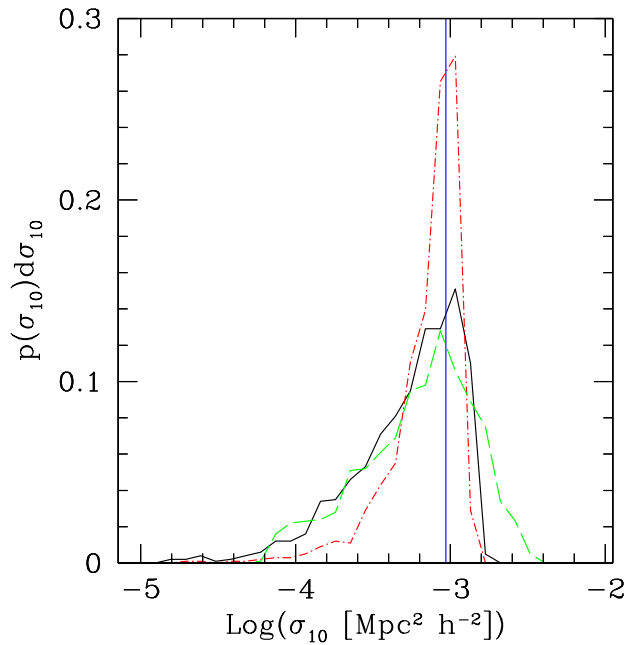}\hfill
\caption{Cross section distributions. The vertical blue line shows
the cross section for arcs with length-to-width ratio $d \ge 10$ computed for
a dark-matter halo of mass $2 \times 10^{15} M_\odot h^{-1}$ at redshift
$z_\mathrm{l} = 0.3$ with
sources at redshift $z_\mathrm{s} = 1$ and lensing-potential ellipticity
$e = 0.3$. The red dot-dashed line is the distribution of the
cross sections caused by the variation of halo concentrations with triaxiality.
The green dashed line includes the ellipticity distribution of projected
triaxial halos, and the black solid line contains both effects.}
\label{fig:effect} 
\end{figure}

Cross sections were computed for each modified halo, using the new
values of the concentration or of the ellipticty, or both.
The three resulting cross-section distributions are shown in
Fig.~\ref{fig:effect}. The variation of the concentration with triaxiality
introduces additional scatter in the cross section (red
dot-dashed line), but significantly less than
the concentration scatter introduced before. The small difference
between the black solid and the green dashed curves in Fig.~\ref{fig:effect}
corroborates this conclusion.
  
The distribution of cross
sections obtained after random projections of triaxial halos is centered on
the cross section for the original halo with fixed ellipticity $e=0.3$,
indicating that this lensing-potential ellipticity is typical.
This confirms the result of \cite{ME03.1}, who found this value by
fitting the deflection angle maps of simulated galaxy clusters (see also
\citealt{ME05.2}). The good
agreement also shows that the reduced concentration of highly triaxial halos
is compensated by the higher ellipticity.

The scatter caused by the ellipticity distribution exceeds that
caused by the variation of the concentration with triaxiality,
but the total scatter in the cross
sections due to halo triaxiality shown in Fig.~\ref{fig:effect} is at most
comparable to that caused by the intrinsic concentration distribution.
Moreover, it does not systematically shift the cross sections towards
higher or lower values, hence leaving unchanged the conclusions of this work.
It should also be noted that these results are expected to hold if more
detailed gas physics (such as cooling and star formation) is included
because it tends to affect the inner slope rather than the ellipticity of
the cluster mass distribution \citep{PU05.1}. 

We have applied the same test to halos of different mass and found very
similar results. The effect of the variation of halo concentrations with
triaxiality on the temperature and luminosity of the X-ray gas is negligibly
small.

\section{Summary and discussion}

We have investigated the effect of the scatter in the relation between
concentration and mass in dark-matter halos on gravitational arc statistics
and X-ray properties of galaxy clusters. 

We have addressed the effect on strong-lensing cross sections of different
implementations of the $c$-$M$ relation proposed in the literature
(\citealt{NA97.1,BU01.1,EK01.1}). We found substantial differences, with the
algorithms by \cite{NA97.1} and \cite{BU01.1} predicting the highest and the
lowest cross sections, respectively. We adopt the algorithm by \cite{EK01.1}
because it needs only one instead of two free parameters, has been shown to be
applicable to cosmological models with dynamical dark energy \citep{DO03.2},
and was found to yield strong-lensing results in good agreement with numerical
simulations. 

This result shows that caution must be applied when modelling galaxy cluster
lenses with NFW density profiles, since different implementations of the
$c$-$M$ relation may yield largely different values for the lensing
efficiency, in particular if axial symmetry is assumed. 

We then used the \cite{EK01.1} algorithm to compute fiducial concentrations
for a sample of $\mathcal{N} = 500$ dark-matter halos with masses between
$10^{14}$ and $2.5 \times 10^{15} M_\odot h^{-1}$ at redshift zero. Each halo
is evolved backwards in time in discrete redshift steps up to a source
redshift randomly drawn for each halo from a parameterisation of the observed
redshift distribution of faint blue galaxies. When the scatter in the
concentration was taken into account, it was drawn from a log-normal
distribution around the fiducial value, with a standard deviation of $\sigma_c
= 0.2$. The effect of cluster mergers on the strong-lensing cross sections was
also included \citep{TO04.1,FE06.1}, although the relative effect of the
concentration scatter is insensitive to mergers. 

The skewness of the log-normal distribution renders concentrations much above
the fiducial value more likely than much below it, thus increasing on average
the strong-lensing cross sections. Thus, the total optical depth, and hence
also the total number of arcs expected on the sky, is increased by up to
$50\%$ by
the concentration scatter. Moreover, the optical depth per unit redshift
displays isolated significant peaks which are due to individual dark-matter
halos with relatively low mass that happen to reach a particularly large
concentration. Such halos can thus be turned into efficient lenses and
contribute strongly to the optical depth because of their high abundance. 

We then used our merger trees to better understand the relationship between
dark-halo concentrations and their lensing efficiency. We found that selecting
halos by concentration yields average cross sections similar to those of the
complete sample. This shows that the higher concentrations of lower-mass halos
compensates for their lower masses in terms of their strong-lensing efficiency
until their caustic curves become too small compared to the sources to produce
large arcs. Massive halos, however, reveal the concentration-dependence of the
strong-lensing cross sections.

Conversely, the median and mean halo concentrations do not increase if the
most efficient lensing halos are selected. However, selecting massive strong
lenses reveals the dependence of the cross sections on the concentration,
yielding median and mean concentrations increasing with the lensing
efficiency. The most massive, strong lenses turn out to be
10-20\% more concentrated than average lensing clusters.

This confirms a bias found earlier in numerically simulated
clusters. \cite{HE07.1} found that strong cluster lenses have three
dimensional concentrations $\approx18\%$ higher than typical clusters with
similar mass. We found that the median concentration is $\approx 12\%$ higher
in halos with very high lensing efficiency compared to average halos with
similar mass, and can grow up to $25\%$ if massive clusters are selected. 

Apart from the qualitative agreement, the quantitative agreement is quite
reassuring especially in view of our different approach of modelling the halo
population and its lensing efficiency semi-analytically compared to fully
numerically. The $12\%$ increase found here is certainly consistent with their
$18\%$ increase because a broader mass selection was applied here. Caution
must thus be applied when extrapolating results on the inner structure of
strongly lensing clusters to the entire cluster population. 

Finally, we performed a similar analysis using the temperature and the
bolometric luminosity of the X-ray emitting intracluster medium instead of the
strong-lensing cross section. We assigned a temperature and an X-ray
luminosity to each dark-matter halo in our sample by extending scaling
relations first derived by \cite{EK98.1}. They maintain the usual scalings $T
\propto [M_{200}h(z)]^{2/3}$ and $L_X \propto M_{200}^{4/3} h(z)^{7/3}$, but
include a dependence on the concentration of the host dark matter halo. 

According to this analysis, there is no concentration bias in
temperature-selected clusters, while a bias similar to strong lensing occurs
for objects selected by their X-ray luminosity, if clusters of similar mass
are selected. In particular, the mean and median concentrations of dark halos
with increasing gas temperature decrease, reflecting that the temperature is
much more sensitive to the halo mass than to its concentration. This result
remains true when the halos are selected by mass. Likewise, dark halos with
increasing X-ray luminosity have virtually unchanged concentrations if no mass
selection is applied. If only massive objects are selected, the dependence of
the bolometric luminosity on the concentration appears. 

It is then an interesting question whether the two concentration biases due to
strong lensing and X-ray luminosity conspire to produce a stronger effect. We
computed the mean and median concentrations of clusters selected for strong
lensing among those already selected for their X-ray luminosity and with mass
larger than $7.5 \times 10^{14} M_\odot h^{-1}$. The further increase in
concentration is very small compared to very X-ray luminous objects only. This
is because selecting massive clusters for their high bolometric X-ray
luminosity, we already select objects with high concentration that are
typically also the most efficient lenses. 

We also checked the effect of halo triaxiality on our results,
which adds scatter to the halo concentrations and projected halo
ellipticities, and, even though the latter is relatively significant, it
leaves the conclusions of our paper unchanged.

These results confirm the general expectation that the gas temperature is more
sensitive to the depth of the overall potential well and thus to the halo mass
than to the internal halo structure. This does not hold true for the
luminosity, which scales with the squared gas density and is thus
substantially more sensitive to structural properties of the halo other than
the mass. Similarly, the lensing efficiency is very sensitive to the details
of the internal structure of the lens, as demonstrated in a variety of studies
\citep{BA95.1,ME03.2,ME03.1,OG03.1,ME07.1}. 

\section*{Acknowledgements}

This work was supported by the Collaborative Research Centre SFB 439 of the
\emph{Deutsche Forschungsgemeinschaft} and by the German Academic Exchange
Service (DAAD) and the Conference of the Rectors of Italian Universities
(CRUI) under the \emph{Vigoni} programme.
Part of the work has been performed under the Project HPC-EUROPA
(RII3-CT-2003-506079) with support by the European Community - Research
Infrastructure Action under the FP6 ``Structuring the European Research Area''
Programme. We thank the anonymous referee for useful comments that helped
improving this paper.

\bibliographystyle{aa}
\bibliography{./master}

\end{document}